# A Feasible Graph Partition Framework for Random Walks Implemented by Parallel Computing in Big Graph

Xiaoming Liu[1], Yadong Zhou[1], Xiaohong Guan[1,2]
[1]Ministry of Education Key Lab for Intelligent Networks and Network Security, Xi'an Jiaotong University, P.R.China
[2]Center for Intelligent and Networked Systems and TNLIST Lab, TsinghuaUniversity, P.R.China
{xmliu, ydzhou, xhguan}@sei.xjtu.edu.cn

## ABSTRACT
Graph partition is a fundamental problem of parallel computing for big graph data. Many graph partition algorithms have been proposed to solve the problem in various applications, such as matrix computations and PageRank, etc., but none has pay attention to random walks. Random walks is a widely used method to explore graph structure in lots of fields. The challenges of graph partition for random walks include the large number of times of communication between partitions, lots of replications of the vertices, unbalanced partition, etc. In this paper, we propose a feasible graph partition framework for random walks implemented by parallel computing in big graph. The framework is based on two optimization functions to reduce the bandwidth, memory and storage cost in the condition that the load balance is guaranteed. In this framework, several greedy graph partition algorithms are proposed. We also propose five metrics from different perspectives to evaluate the performance of these algorithms. By running the algorithms on the big graph data set of real world, the experimental results show that these algorithms in the framework are capable of solving the problem of graph partition for random walks for different needs, e.g. the best result is improved more than 70 times in reducing the times of communication.

## Categories and Subject Descriptors
G.2.2 [Mathematics of Computing]: Graph Theory – Network problems; G.3 [Mathematics of Computing]: Probability and Statistics - Markov processes;

## 1. INTRODUCTION
With the development of web 2.0, the online social networks members are increasing sharply. Faced with the challenge of big graph data, more and more researchers pay great amount of attention to parallel graph computing. Up to now, many parallel graph computing systems are constructed, e.g. GraphX [1], PowerGraph [2], Pregel [3]. Graph partition algorithm [4] is used for cutting the graph into subgraphs called partitions. It is fundamental problem of the parallel computing on graph, but usually difficult for some complicated graph analysis methods, i.e., random walks. The theory of random walks [5] is employed for exploring the structure and function of graphs in various fields, such as graph sampling and estimating [6], community detection [7], etc. When the walker walks in partitions, it may jump from one partition to another one frequently. This leads to communication between partitions. So the partitions are not independent, which is contrary to requirements in traditional parallel computing. Copious communication between partitions will cost lots of bandwidth and computing resource. The large number of the replications of the vertices would increases the expense of storage and memory. The unbalanced partition leads to unbalanced load and has adverse effect on the efficiency of parallel computing. A good partition has signification in parallel computing on graph. It should make the partitions occupy less storage and memory, reduce the cost of bandwidth and computing resource, and ensure the load balance of parallel graph computing.

Graph partition for parallel graph computing is a crucial problem. Many outstanding works have been done in this field. These works can be divided into two categories, i.e., theoretical analysis and practical application. In the first category, the researchers [8] [9] [10] [11] try to find the nature of graph partition and provide methodologies for applications. These methodologies have perfect theoretical analysis. But they are hardly applied to graph partition for parallel graph computing for limitations, such as worse time complexity, the lack of considering requirements of practical application, etc. One example is that Lang [8] proposes a few algorithms to find the good balanced cuts in power law graph, but he doesn't consider the occupation of storage and memory, and the number of times of communication, actually. So even in some parallel graph computing systems, the graphs are partitioned randomly [2], which leads to poor performance. In the second category, the researchers focus on the practical application for graph partition for parallel computing. They propose many simple but effective algorithms. Instead of proposing a general algorithm, the researchers makes their algorithms only work for one certain problem, such as matrix computation [12], triangle listing [13], PageRank [14], etc. In this situation, the algorithm can work well on the certain problem in the application of parallel graph computing. Random walks is a widely used method in various fields. To best of our knowledge, no one has done the work of graph partition for random walks.

In this paper, instead of proposing a general algorithm, we also focus on one special case, i.e., the graph partition algorithm for random walks. In the process of random walks on partitions implemented by parallel computing, the cost of bandwidth depends on the number of times of communication, i.e., how many times the walker jumps from one partition to another one. The occupation of memory and storage is determined by the number of replications of the vertices. Intuitively, a small number of edges between partitions would reduce the cost of bandwidth; the few replications of the vertices could lower the occupation of memory and storage. By analyzing the process of random walks on partitions, we give two optimization functions. The two optimization functions are used for lowering the number of replications of the vertices and reducing the number of times of communication between partitions. The constraint makes sure that the partition are well-distributed as possible as they can. The two optimization functions and their constraints are the foundation of the framework for graph partition for random walks. In this framework, we propose four graph partition algorithms. We also list their advantages and shortcomings. In these algorithms, we try to achieve the targets: 1) the number of the vertices in each partition should be nearly the same; 2) the number of the cut vertices should be as few as possible; 3) the cut vertices should have relative small degrees; 4) the number of the edges between partitions should be few; 5) the running time should be acceptable; 6) the graph in partition should be connected. We also propose five metrics to evaluate the performance of these graph partition algorithms. We apply our algorithms in graph data of real world, and experimental results show the different advantages of our algorithms.

The contributions of our work are listed below.

1) We analyze the process of random walks on partitions, and transform the graph partition for random walks problem into optimization problem. Two optimization functions are presented and proved to be able to reduce the number of replications of the vertices and the number of the times of communication;

2) Based on the two optimization functions, we propose a feasible graph partition framework for random walks implemented by parallel computing. In this framework, four greedy graph partition algorithms are listed. We analyze their advantages and shortcomings which are shown in experimental results. Because graph partition is the data preprocessing of parallel graph computing, its time complexity should not be too large, otherwise it will cost more time than the computing without parallelism. In our algorithms, the worst time complexity of these algorithms is better than $O(n^2)$.

3) We propose an integrated metric system for the problem of graph partition for random walks. The system is composed by five metrics. Experimental results demonstrate that these metrics can reflect the performance of these algorithms strictly.

The rest of this paper is organized as follows. Section 2 presents some notions and preliminaries. Section 3 is mathematical analysis for optimization functions and integrated metric system. Section 4 is the sketch of the greedy algorithms of graph partition for random walks. Section 5 gives the experimental results. Section 6 is about the related work. Section 7 summarizes the conclusion and future work.

## 2. PRELIMINARIES AND NOTIONS
Since we try to solve the problem of graph partition for random walks implemented by parallel computing, the preliminaries and notions of random walks will be introduced first.

## 2.1 Preliminaries on Random Walks
A network is denoted by an unweighted undirected graph $G = (V, E)$, where $V = \{v_1, v_2, \ldots, v_n\}$ is the vertex set and $E = \{e_{ij}, \ldots, e_{lf}\}$ is the set of edges. $e_{ij} = (v_i, v_j)$ and $e_{ji} = (v_j, v_i)$ present the same edge. Let $n = |V|$ and $m = |E|$. The degree of the vertex $v_i$ is denoted by $d(i)$, where $1 \leq i \leq n$.

The *adjacency matrix* of graph $G$ is denoted by a $n \times n$ matrix $A$. Namely, if $v_i$ and $v_j$ are connected, $A_{ij} = A_{ji} = 1$, and $A_{ij} = A_{ji} = 0$ otherwise.

If the walker chooses a vertex $v_i$ among the neighbors of the current vertex $v_j$ randomly and uniformly, the process of random walks on graph $G$ is a *Markov process* [15], and the *state space* is the vertex set of graph $G$. The *transition matrix* is defined as $P$, so

$$P = D^{-1}A \qquad (1)$$

where $D$ is the *diagonal matrix* of the vertex degrees.

$$D_{ij} = \begin{cases} d(i), & i = j \\ 0, & i \neq j \end{cases} \qquad (2)$$

At each step, the *transition probability* from $v_i$ to $v_j$ is

$$P_{ij} = \frac{A_{ij}}{d(i)} \qquad (3)$$

The *transition probability* from $v_i$ to $v_j$ through walking $t$ steps randomly is denoted by $P_{ij}^{(t)}$. When the number of steps $t$ tends towards infinity, the probability is [16]

$$\lim_{t \to \infty} P_{ij}^{(t)} = \frac{d(j)}{\sum_f d(f)} \qquad (4)$$

The *transition probability* is independent of the start vertex $v_i$, and just depends on the degree of end vertex $v_j$.

The sequence of vertices visited by the walker is a *Markov chain*, and the *stationary probability distributions* of Markov chain follows

$$\pi P = \pi \qquad (5)$$

where $\pi = \{\pi_1, \pi_2, \ldots \pi_n\}$ is the *stationary probability distribution*. Because the connected undirected graph $G$ of complex network with finite vertices is not a bipartite graph, the *stationary probability distribution* of $v_j$ follows

$$\pi_j = \lim_{t \to \infty} P_{ij}^{(t)} \qquad (6)$$

When the probability distribution reaches stationary, known from (4) and (6), the probability of traversing any edge $e = (v_i, v_j)$ of $G$ is

$$p(e) = \pi_j \cdot \frac{1}{d(j)} + \pi_i \cdot \frac{1}{d(i)} = \frac{2}{\sum_f d(f)} \qquad (7)$$

## 3. MATHEMATICAL ANALYSIS

Random walks is a widely used method for sampling and estimating networks, detecting the structure of the network, etc. Graph partition is also one of the most important issues in the field of parallel graph computing. In this section, we focus on the analysis of graph partition for random walks, and propose the metrics for graph partition result.

### 3.1 Mathematical Analysis of Graph Partition for Random Walks

The number of times of communication and the number of replications of the vertices are two key points for parallel computing on partitioned graph. The fewer number of times of communication means the less occupation of bandwidth, and the more times of the vertices to be replicated imply the larger requirement of the memory and storage of compute nodes. By analyzing the process of random walks on partitioned graph, we give the optimization functions to reduce the number of communication and replications of the vertices.

All the analysis is based on the situation that the walker has reached the *stationary probability distribution*. Assume that the graph is partitioned into $k$ parts, and the set of the partitions is denoted by

$$Pas = \{Pa_1, Pa_2, \ldots, Pa_k\} \tag{8}$$

In order reduce the occupation of the memory and storage, the number of the replications of the vertices should be minimum. So the optimization problem is

$$\min_{NR} \frac{1}{|V|} \sum_{v \in V} NR(v)$$

$$s.t. \max_m \{Pa \in Pas \mid Ne(Pa) = m\} \leq \lambda \frac{|E|}{k} \tag{9}$$

where $NR(v)$ denotes the number of replications of vertex $v$, $Ne(Pa)$ donates the number of the edges in partition $Pa$, and . $\lambda$ is the unbalance parameter for the partition.

From (9) we can find that the optimization objective function is set up for that the number of replications of the vertices reaches the fewest, and the constraint makes sure that the partitions are well-distributed as possible as they can.

The analysis about reducing the number of times of communication is given below.

Suppose that $v_j$ is partitioned into $s$ parts by vertex-cut method [17]. $d_{in}(Pa_i v_j)$ denotes the number of the edges of $v_j$ in $Pa_i$, where $Pa_i \in Pas$, and $d_{out}(Pa_i v_j)$ denotes the number of the edges of $v_j$ out of $Pa_i$. When the walker reaches $v_j$, the probability of the walker jumping out of $Pa_i$ is

$$probout(Pa_i v_j) = \frac{d_{out}(Pa_i v_j)}{d(j)} \tag{10}$$

where $d(j)$ is the degree of $v_j$, and $d(j)$ also meets that

$$d_{out}(Pa_i v_j) = d(j) - d_{in}(Pa_i v_j) \tag{11}$$

Then, the probability that the walker jumps out of one partition from $v_j$ is

$$prob_{out}(v_j) = \sum_{Pa_i \in Pas} \frac{d(j)}{\sum_f d(f)} \frac{d_{in}(Pa_i v_j)}{d(j)} \frac{d_{out}(Pa_i v_j)}{d(j)}$$

$$= \sum_{Pa_i \in Pas} \frac{d_{in}(Pa_i v_j)}{\sum_f d(f)} \frac{d_{out}(Pa_i v_j)}{d(j)} \tag{12}$$

From (12), we can gain the probability that the walker jumps out of one partition for each step is

$$prob_{out} = \frac{1}{\sum_f d(f)} \sum_{v_j \in V} \sum_{Pa_i \in Pas} \left( d_{in}(Pa_i v_j) - \frac{d_{in}^2(Pa_i v_j)}{d(j)} \right) \tag{13}$$

Then, the expectation of the times of communication for each step is

$$E(C) = \frac{1}{\sum_f d(f)} \sum_{v_j \in V} \sum_{Pa_i \in Ps} \left( d_{in}(Pa_i v_j) - \frac{d_{in}^2(Pa_i v_j)}{d(j)} \right) \tag{14}$$

From (14), the optimization problem of reducing the times of communication is gained.

$$\min_{Pas} \frac{1}{|Pas| \sum_f d(f)} \sum_{v_j \in V} \sum_{Pa_i \in Pas} \left( d_{in}(Pa_i v_j) - \frac{d_{in}^2(Pa_i v_j)}{d(j)} \right)$$

$$s.t. \max \{V_i \in Vs \mid |V_i| = n_i\} \leq \lambda \frac{n}{|Pas|} \tag{15}$$

where the set $Vs = \{V_1, V_2, \ldots, V_k\}$ is composed of the vertices in each partition.

The optimization objective function in (15) is used to optimize the number of times of communication, and the constraint also makes sure that the partitions are balanced.

The solution of two optimization functions (9), (15) provides the best partitions for parallel computing of random walks. But this is a multiobject optimization problem (MOP) [18] for graph modularity. The time cost of finding the best solution is very expensive for MOP, or even the best solution cannot be found. And modularity optimization is also an NP-complete problem [19]. In the big data era, any algorithm with high time complexity is often undesirable. Besides, because graph partition is the data preprocessing of parallel computing, its time complexity should not be too large, or it will cost more time than the computing without parallelism. Then, instead of finding the best solution of the MOP, we propose a feasible graph partition framework for random walks implemented by parallel computing in big graph through analyzing (9) and (15). In this framework, several greedy algorithms are proposed, which are presented in section 4.

### 3.2 The Metric Analysis for Graph Partition

In this part, we propose five metrics to evaluate the performance of the graph partition algorithm. The five metrics include modularity, balance, running time, connection, and vertex-cut. Running time means the time cost of the algorithm to get the partitions. The connection means that whether the partition is connected or not. These two metrics can be easily gotten by the results of graph partition. Less running time is better. Since graph data is dependent, the connected partition is beneficial to parallel graph computing. Next, we give detailed analysis for modularity, balance and vertex-cut.

### 3.2.1 Modularity
From (15) we can find that the situation that the more edges in partition is beneficial to reduce the number of times of communication. As for a good graph partition, they should have most of the edges inside partitions, but few edges between them. This is very similar to the definition of community [20] in social networks.

For one partition $Pa_i$ in $Pas$, the true probability of an edge in $Pa_i$ is

$$tp = \frac{\sum_{v_j \in Pa_i} d_{in}(Pa_i v_j)}{2m} \quad (16)$$

If the edge is selected randomly in graph $G$, the probability that at least one end in partition $Pa_i$ is $\sum_{v_j \in Pa_i} d(i)/2m$. So both ends in $Pa_i$ is $(\sum_{v_j \in Pa_i} d(i)/2m) \cdot (\sum_{v_h \in Pa_i} d(h)/2m)$. So the expected probability of an edge in $Pa_i$ is

$$ep = \frac{\sum_{v_j \in Pa_i} d(j)}{2m} \cdot \frac{\sum_{v_h \in Pa_i} d(h)}{2m} \quad (17)$$

So for $Pa_i$, the improvement is

$$imp = tp - ep \quad (18)$$

This just satisfies correlation analysis of *Leverage* [21]. Lian Duan et al [22] give the similar analysis for community detection.

From (16), (17), we can get that $ep \geq tp^2$. Then,

$$imp \leq tp - tp^2 \quad (19)$$

Since $tp \in [0,1]$, the upper bound of the improvements for one partition is 0.25.

For all the partitions, the total improvement is

$$\sum_{i=1}^{k} imp_i = \sum_{i=1}^{k} \left[ \frac{\sum_{v_j \in Pa_i} d_{in}(Pa_i v_j)}{2m} - \frac{\sum_{v_j \in Pa_i} d(j)}{2m} \cdot \frac{\sum_{v_h \in Pa_i} d(h)}{2m} \right]$$
$$= \frac{1}{2m} \sum_{v_j, v_h \in G} \left( A_{ij} - \frac{d(j)d(h)}{2m} \right) \cdot \delta(v_j, v_h) \quad (20)$$

$\delta(v_j, v_h)$ is 1 if $v_j$ and $v_h$ are in the same partition, otherwise 0. The total improvement is the same to the definition of modularity [23], and its value ranges from 0 to 1.

### 3.2.2 Balance
The balance of the graph partition can guarantee that the computing resource is fully utilized and its load balance. Here we just consider the vertex distribution in the partitions. The expected number of the nodes in partition is $N/k$, where $N$ is total vertices in partitions. $N_i$ is the number of the vertices in partition $Pa_i$. The balance variance for $Pa_i$ is defined as

$$var_i = \frac{\left(N_i - N/k\right)^2}{N^2} \quad (21)$$

So $var_i$ ranges from 0 to $\left(\frac{k-1}{k}\right)^2$. For the total balance variance for all the partitions is

$$var = \frac{\sum_{i=1}^{k}\left(N_i - \frac{N}{k}\right)^2}{N^2}$$
$$\leq \left( \frac{N^2}{k^2}(k-1) + \frac{(k-1)^2}{k} N^2 \right) / N^2 \quad (22)$$
$$= \frac{k-1}{k}$$

So $var \in \left[0, \frac{k-1}{k}\right]$. The smaller value is better for parallel computing load balance [24].

### 3.2.3 Vertex-cut metric
The number of replications of the vertices determines the cost of the storage and memory. Here we introduce the metric for vertex-cut to evaluate the performance of the algorithm. $NC_i$ is the number of the cut vertices in $Pa_i$. The true probability that the vertex is cut vertex in in $Pa_i$ is

$$tp = \frac{NC_i}{N} \quad (23)$$

If we select one vertex randomly, the expected probability that the vertex is a cut vertex in $Pa_i$ is

$$ep = \frac{N_i}{N} \sum_{v_j \in Pa_i} \frac{1}{N_i} \left[ 1 - \left(\frac{N_i}{N}\right)^{d(j)} \right]$$
$$= \frac{1}{N} \sum_{v_j \in Pa_i} \left[ 1 - \left(\frac{N_i}{N}\right)^{d(j)} \right] \quad (24)$$

The improvement is

$$impc_i = tp - ep \quad (25)$$

If $impc_i < 0$, it means that the algorithm can reduce the number of replications of the vertices, compared with random hash; otherwise, it means the algorithm even performs worse than random hash.

The total improvement is

$$\sum_{i=1}^{k} impc_i = \sum_{i=1}^{k} \left\{ \frac{NC_i}{N} - \frac{1}{N} \sum_{v_j \in Pa_i} \left[ 1 - \left(\frac{N_i}{N}\right)^{d(j)} \right] \right\} \quad (26)$$

The total improvement ranges from -1 to 1.

## 4. THE GREEDY ALGORITHMS
In this section, we focus on the analysis of the two optimization functions and explore the rules of the greedy graph partition algorithms for random walks.

### 4.1 The Framework of the Algorithms
By surveying these two optimization functions, we discover these features:

1) The fewer vertices to be cut are, the better for decreasing the number of replications of the vertices and number of times of communication it is, on the premise of meeting the constraints;

2) For each cut vertex, the fewer replications the vertex has, the better for reducing the number of replications of the vertices it is;

3) For each cut vertex, the average distribution of its edges in the partitions is the worst case for reducing number of times of communication;

4) For each cut vertex, the larger degree the vertex has, the worse for reducing the number of replications of the vertices and number of times of communication it is.

These four features are the fundamental of greedy algorithms of graph partition for random walks. For the sake of decreasing running time, the algorithms may do not fit the four rules listed above perfectly. In the algorithms, we try to achieve the following targets:

1) The number of vertices in each partition is nearly same;
2) The cut vertices should have relative small degree;
3) The number of cut vertices had better be as few as possible;
4) Every two vertices in each partition are connected.

So we can get the framework of the graph partition for random walks.

1) The large degree vertex should be merged into partition firstly;
2) The large degree vertex and its neighbors had better be in the same partition;
3) There should have rules to ensure the balance of the partitions;

By analyzing the features of random walks on partitions, we has transformed graph partition for random walks into a very simple problem. Based on the framework, four simple and feasible algorithms are proposed.

## 4.2 The Sketch of the Greedy Algorithms
In this part, we introduce the four algorithms from the simple to complex. The four algorithms are simple but effective for graph partition (GP) for random walks.

### 4.2.1 Breadth first search
From the framework we can get that the large degree vertex and its neighbors had better be in same partition. The first idea comes into our brain is breadth first search (*BFS*) which can achieve the target. Based on breadth first research, we propose a greedy algorithm.

Here we introduce the main idea of the algorithm. First, we take the $k$ unconnected vertices with the $k$ top large degree as the initial vertices of the $k$ partitions, respectively. Afterwards, we do *BFS* for each partition to find the neighbors of the partition and merge them into partition. When no vertex can be merged any more, the algorithm stops.

**Algorithm 1** The *BFS* GP Algorithm

**Input** $k, G$

**Output** *Pas*

1. Find the $k$ unconnected vertices with top large degree;
2. Initial each partition by merging one of $k$ vertices;
3. **while** |V| !=0 **do**
4.   **for** $i$ **in** range($k$) **do**
5.     Merge the neighbors which has not been in Partitions into $Pa_i$ and remove these vertices from $V$;
6.   **end**
7. **end**
8. **Return** *Pas*;

The advantages of the algorithm are listed. The running time will be short; because all the vertex's neighbors are merged into one partition, the modularity of the partitions should be large; the vertices with large degree are also merged firstly with other ones, which are beneficial to modularity; each partition is connected; The shortcomings are the balance of partition may be not good enough, because in each around different partition may merge different number of vertices.

### 4.2.2 Large degree first search
Similar to breadth first research, another strategy also works. It is large degree first search (*LDFS*) which meets the demands of the large degree vertex should be merged into partition firstly.

Here we introduce the main idea of the algorithm. First, we take the $k$ unconnected vertices with the $k$ top large degree as the initial vertices of the $k$ partitions, respectively. Then find the vertex with largest degree which has neighbor in partition, the vertex will be merged into the partition. If the vertex has more than one options, the vertex will be merge into the partition which has its neighbor with largest degree. When there is no vertex to process, the algorithm stops.

**Algorithm 2** The *LDFS* GP Algorithm

**Input** $k, G$

**Output** *Pas*

1. Find the $k$ unconnected vertices with top large degree;
2. Initial each partition by merging one of $k$ vertices;
3. **while** |V| !=0 **do**
4.   **for** $i$ **in** range(|V|) **do**
5.     **If** $V[i]$ has neighbors with in partitions **do**
6.       Merge $V[i]$ in the partition including the vertex with largest degree in neighbors, and remove $V[i]$ from $V$;
7.     **end**
8.   **end**
9. **end**
10. **Return** *Pas*;

The above algorithms have similar advantages and shortcomings. Because in every round each partition only merge one vertex, it may broke the closely connected subgraphs, which may decrease modularity. So the modularity may be reduced.

### 4.2.3 Balance-Combine BFS and LDFS

The shortcoming of the above two algorithms is potential unbalanced partition. In this algorithm we start considering the balance of the partition. The algorithm combines *BFS* and *LDFS*.

Here we introduce the main idea of the algorithm. First, we take the $k$ unconnected vertices with the $k$ top large degree as the initial vertices of the $k$ partitions, respectively. Then in every round, each partition merges one vertex with the largest degree which is not in other partitions in its neighbors. When all the vertices are merged into partitions, the algorithm stops.

**Algorithm 3** The *balance* GP Algorithm

**Input** $k$, $G$

**Output** Pas

1. Find the $k$ unconnected vertices with top large degree;
2. Initial each partition by merging one of $k$ vertices;
3. **while** $|V|$ !=0 **do**
4.   **for** $i$ **in** range($k$) **do**
5.     Merge one neighbor with largest degree which has not been in Partitions into $Pa_i$ and remove the vertex from $V$;
6.   **end**
7. **end**
8. **Return** *Pas*;

The running time of the algorithm will be much longer than *BFS* and *LDFS*, but the balance of the partitions will be improved significantly. The modularity may be reduced for the rules of balance.

### 4.2.4 Vertex-cut based on balance

Based on previous algorithm (*Balance*), another algorithm is proposed by cutting vertex. This algorithm will reduce the number of communication.

Here we introduce the main idea of the algorithm. First, we take the $k$ unconnected vertices with the $k$ top large degree as the initial vertices of the $k$ partitions, respectively. Then in every round, each partition merges one vertex with the largest degree in its neighbors. If the vertex has been in other partitions, the vertex will be cut into two parts. The vertex cut example can be seen in Fig. 1. In this algorithm, the key point is how to manage the neighbors of the partitions effectively.

Here is the rules to manage the neighbors of $Pa_i$ in the partitions:

1) If the vertex has not been in other partitions, all the neighbors of the vertex will be merged into the neighbor set of the partition $Pa_i$;

2) If the vertex has been in other partitions, all of the neighbors of the vertex which are not in any partition are merged in to the neighbor set of the partition $Pa_i$, and we need to check which partitions the vertex has been in, find out the neighbors of the vertex which are in which are in $Pa_i$, and remove these neighbors from the neighbor sets of these partitions.

In complex networks, like social networks, members are connected so closely. In closely connected graph, the vertex is easily cut. These rules ensure that the number of the cut vertices is not too large, or in each partition, the number of the vertices is nearly same to $n$.

**Algorithm 4** The *vertex-cut* GP Algorithm

**Input** $k$, $G$

**Output** Pas

1. Find the $k$ unconnected vertices with top large degree;
2. Initial each partition by merging one of $k$ vertices;
3. **while** $|V|$ !=0 **do**
4.   **for** $i$ **in** range($k$) **do**
5.     Find the neighbor with largest degree;
6.     **if** the neighbor is not in any partition **do**
7.       Merge the vertex into $Pa_i$;
8.     **else do**
9.       Cut the vertex and merge the vertex into $Pa_i$ and reorganize the neighbors of the partitions by rules;
10.    **end**
11.  **end**
12. **end**
13. **Return** *Pas*;

The running time will be much longer than the previous ones for the reason that managing the neighbors of the partitions is really complex. But compared with *Balance*, this algorithm has large modularity by the replications of the vertices.

## 4.3 Implement Random Walks on Partitions by Parallel Computing

In order to implement the walker walk on the partitions by parallel computing, we design three tables to represent the partitions and their relationships. An example of the tables is given in Fig. 1. Fig. 1a is a graph which is partitioned into two parts by cutting vertex $v_3$. Fig. 1b shows the two partitions. Fig. 1c and Fig. 1d are two tables of partition1 and partition2 respectively which are called vertex-neighbor tables. *NoP* denotes the neighbors of vertex in this partition, and *NoOP* presents the left neighbors of vertex in other partitions. Fig. 1e and Fig. 1f show the relationships between vertices and partitions, and the table is called vertex-partition table.

When the random walks on partitions begins, the start vertex is selected randomly and uniformly. We can know the ID of the partition which the start vertex belongs to by vertex-partition table. If there are more than one partitions the start vertex belongs to, e.g., $partition_1$, $partition_2$, ... , $partition_x$, and the numbers of edges in these partitions are $m_1$, $m_2$, ... , $m_x$ respectively, the probability of choosing $partition_f$ as the graph data of the job is $m_f / \sum_i m_i$. During the process of random walks, whether or not the vertex belongs to one partition can be known by *NoOS* in vertex-neighbor table (or from vertex-partition talbe). If the neighbors of the current vertex are only in one partition, the random walks is just carrying on as the graph not being partitioned; otherwise, the next vertex is selected from vertices in *NoP* and *NoOP* of the current vertex. If the selected vertex is in other partitions, it will be processed by the same strategy of the start vertex. So we can find that the result of random walks on partitioned graph has no difference from the walks on the graph without partitioning. Therefore, we can employ multiple independent random walks on partitions by parallel computing.

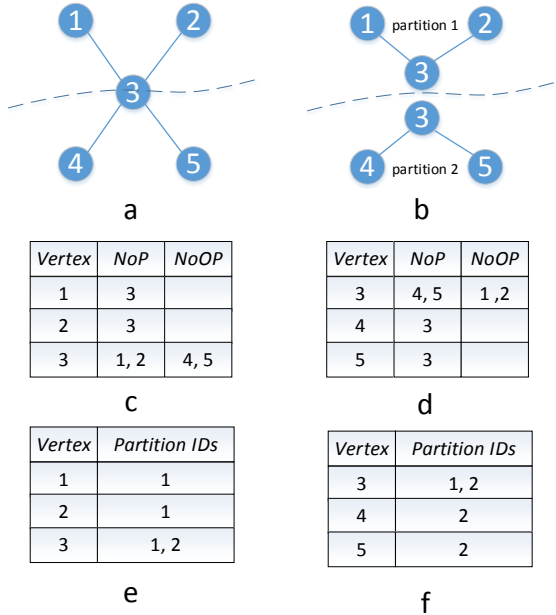

**Fig. 1 An example of graph partition by vertex-cut and the related tables created for the parallel computing of random walks.**

## 5. EXPERIMENTAL RESULTS

We test our algorithm on the graph in the real world. The graph data is the network of the friendship in Facebook [25]. The data of Facebook is crawled with the strategy of *BFS*. Social networks is one special case of complex networks, it is very difficult graph partition problem for the large data set, an uneven degree distribution. The number of the partitions is set as 10. We implement the experiments on two different scales of data, i.e., the graph with 100001 vertices and the graph with 1000001 vertices.

### 5.1 The Experimental Data

Before the experimental results are present, we briefly introduce the experimental data. In the raw data of Facebook, there are several abnormal vertices which has no neighbors in graph but they appear in the neighbors of other vertices. We fix it by add one vertex to the graph and take the vertex as the neighbor of these vertices. That's why there always have one vertex seems to be redundant. Fig. 2 and Fig. 3 are the degree distributions of the two graphs.

From Fig. 2 and Fig. 3, we can see the uneven degree distribution of complex networks. The few vertices with large degree almost connect to all the others vertices. This closely connected vertices are very hard to partition. If our algorithms work well in complex networks, they should have the ability to solve the graph partition problem in other types of networks.

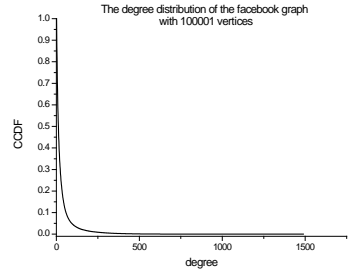

**Fig. 2. The degree distribution of the graph with 100001 vertices.**

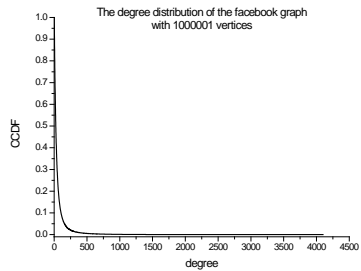

**Fig. 3. The degree distribution of the graph with 1000001 vertices.**

### 5.2 The Experimental Results

In this part, we introduce the experimental results on the two graphs. We analyze the result by the five metrics proposed in this paper. The five metrics include modularity, balance, running time, connection and the vertex-cut metric. We also running random walks on the partitions which are gotten by the algorithms.

Table 1 is the comparison of five algorithms on the graph with 100001 vertices. The five algorithms include the four algorithms proposed in this paper and the random hash algorithm. In random hash algorithm, the vertices are assigned into each partition in the same probability. In Table 1, second is presented by '*s*', and 'YES' means that each partition is a connected subgraph of the graph, and 'null' means that there is no vertex-cut in the processing of the algorithm so the metric for cut vertex is null.

**Table 1. The algorithm comparison**

| Algorithm | Modularity | Balance | Time | Connection | Vertex-cut |
|---|---|---|---|---|---|
| BFS | 0.64 | 0.12 | 93.91s | YES | null |
| LDFS | 0.49 | 0.26 | 2.68s | YES | null |
| Balance | 0.21 | 0.00 | 431.50s | YES | null |
| Vertex-cut | 0.37 | 0.00 | 946.17s | YES | -0.99 |
| Random hash | 0.00 | 0.00 | 3.70s | NO | null |

The *BFS* graph partition algorithm has the largest modularity, which is an advantage, as we analyze in section 4. The large modularity indicates that it will work well in reducing the number of times of communication, which will be shown in the test of implementing random walks on these partitions. The running time is also acceptable. But it has no satisfactory result in the balance of the partitions. When the data is becoming larger and larger, the unbalanced partitions may be a potential trouble for the parallel graph computing, and we should add new rules to deal with this trouble. The unbalanced partitions are caused by the structure of the complex networks. Compared with *BFS*, *LDFS* has clear advantage in running time, but the decreased modularity will lead the increase of communication and the quality in balance becomes worse which will aggravate the potential trouble in *BFS*. The *balance* algorithm solves the problem of unbalanced partition, but the running time increase sharply. The *vertex-cut* algorithm has good modularity than *balance* because it adopts *vertex-cut* strategy. The replications of the vertices make the vertices connect more closely. This algorithm also works well in the quality of balance, but the running time is also a challenge for it. We can see the value of vertex-cut metric is very well (the best value is -1), which indicates that the algorithm based on our analysis reduce the number of replications of the vertices sharply. Although the replications of the vertices cost more storage and memory, it can reduce the number of communication. The number of the vertices of *vertex-cut* algorithm in each partition is show in Table 2. For the *random hash* algorithm, it is simple and has advantages in running time and balance. But the modularity is 0.00 which means that the walker very easily walks out one partition. The partition is also unconnected. The number of the subgraphs in each partition is shown in Table 3. These unconnected subgraphs in partitions make the walker more easily jump out the partitions and walks inefficiently.

Table 2. The number of the vertices in each partition

| Partition ID | The number of vertices | Partition ID | The number of vertices |
|---|---|---|---|
| 1 | 34963 | 6 | 34388 |
| 2 | 34965 | 7 | 39500 |
| 3 | 34443 | 8 | 35017 |
| 4 | 34384 | 9 | 34506 |
| 5 | 34707 | 10 | 34418 |

From Table 2, we can see that there are 251290 replications of the vertices. This is because the close connected vertices in social networks. In social networks, the average distance form one person to another one is 6. If we partition the graph of social networks into 10 partitions, almost each two partition are connected by many edges. That's why so many vertex are cut. And maybe for social networks, the algorithms based on *edge-cut* [1] are much better than the ones based on *vertex-cut*, considering the cost of the storage and memory, even if we have not proved it now.

Table 3. The number of the subgraphs in each partition

| Partition ID | The number of subgraphs | Partition ID | The number of subgraphs |
|---|---|---|---|
| 1 | 755 | 6 | 711 |
| 2 | 729 | 7 | 761 |
| 3 | 703 | 8 | 714 |
| 4 | 759 | 9 | 699 |
| 5 | 705 | 10 | 734 |

From Table 3 we can see that there are too many subgraphs in each partition. Actually, there are only nearly 10000 vertices in each partition. This means that the vertices are distributed desultorily in partitions, which has no beneficial effects to random walks.

Table 4 is comparison of five algorithms on the graph with 1000001 vertices. Because the running time of the vertex-cut algorithm is more than 10 days and the result is still not output, so the metrics for it are all nulls.

Table 4. The algorithm comparison

| Algorithm | Modularity | Balance | Time | Connection | Vertex-cut |
|---|---|---|---|---|---|
| BFS | 0.43 | 0.19 | 6625.63s | YES | null |
| LDFS | 0.30 | 0.17 | 242.06s | YES | null |
| Balance | 0.03 | 0.00 | 280045.58s | YES | null |
| Vertex-cut | null | null | null | null | null |
| Random hash | 0.00 | 0.00 | 68.81 | NO | null |

Compared to Table 1, these algorithms have similar trend in these metrics. The values of their modularity decrease generally. Because the data of Facebook is crawled by breadth first strategy, so the vertices are connected sparsely in last layers of *BFS*. The total number of the member in the raw data set is about 1100000 vertices, then the graph is not connected so closely when the number of the vertices reaches one million. The running time also degreases drastically, this will be a challenge in the larger data set. Actually, the time complexity is low and strategy is very simple now. It indicates that one simple and effective algorithm has the great significance in the big data era.

Table 5 is the number of subgraphs in each partition for *random hash* algorithm. The data set is the graph with 1000001 vertices.

Table 5. The number of the subgraphs in each partition

| Partition ID | The number of subgraphs | Partition ID | The number of subgraphs |
|---|---|---|---|
| 1 | 2519 | 6 | 2536 |
| 2 | 2495 | 7 | 2531 |
| 3 | 2435 | 8 | 2499 |
| 4 | 2520 | 9 | 2516 |
| 5 | 2509 | 10 | 2476 |

From Table 5, when the number of the vertices reaches millions, the number of the subgraphs in each partition is still too many. This indicates the bad performance in the process of random walks on these partitions.

In order to give more intuitive comparison of these algorithms, we implement random walks on these partitions. We count the number of the steps the walker walks around in one partition. When the walker jumps out and walks into another partition, the counter restarts. The larger number of the steps in one partition indicates the fewer times of communication. Fig. 4 is the CCDF result on the graph with 1000001 vertices.

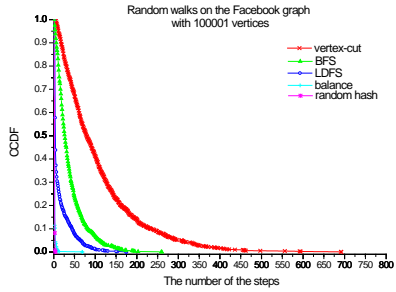

**Fig. 4. The statistic of the steps of random walks on partitions.**

From Fig. 4, we can see that the partitions gotten by *random hash* algorithm performs so badly in the process of random walks. Every time the walker just walks several steps in each partition before jumping out. The algorithms with large modularity perform well in this comparison. This indicates incontestably that the modularity metric can present the performance of the algorithm in reducing communication. The four algorithms proposed in this paper have clear advantage in reducing the communication. Specially, the *vertex-cut* algorithm has the largest advantage in this comparison at the cost of occupying more storage, memory and running time.

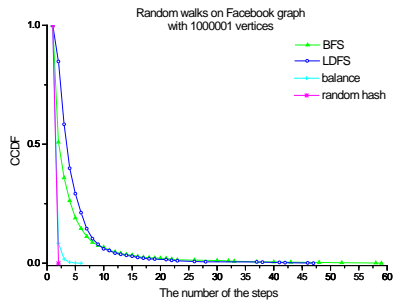

**Fig. 5. The statistic of the steps of random walks on partitions.**

Fig. 5 is the CCDF result on the graph with 1000001 vertices. Compared with the result in Fig. 4, the result in Fig. 5 has similar trend. The performance of all the algorithm becomes worse for the decreasing of modularity of the partitions. The reason for why the modularity decreases is shown in previous analysis.

## 6. RELATED WORK

Graph partition for big graph is an important problem for parallel graph computing. A good partition algorithm has signification to parallel computing on big graph. It can make the partitions occupy less storage and memory, reduce the cost of bandwidth and computing resource, and ensure the load balance of parallel graph computing. Many works have been in this field of graph partition. They can be classified into two categories. One is about theoretical analysis, and another one focuses on practical application.

**Theoretical analysis** These researchers give plenty of perfect theoretical analysis and proofs in their work. They try to find the nature of graph partition and provide general methodologies for different applications in the parallel graph computing. But few ones can really be implemented for limitations, such as worse time complexity, the lack of considering the requirements of practical application, etc. Kevin Lang [8] proposes a few algorithms to find the good balanced cuts in power law graph, but he doesn't consider the occupation of storage and memory, and the number of the times of communication, actually. Chris H.Q. Ding et al. [9] provide a min-max algorithm for graph partition and data cluster with an objective function. The objective function follows the min-max clustering principle. The partition result is obtained by searching the solution of the objective function. The time complexity is a challenge for the algorithm in the big data era. Xing, E. P et al. [10] propose graph partition strategies for generalized mean field inference by MinCut. Abou-Rjeili, A. et al. [11] propose several multilevel algorithms to partition power-law graphs by graph spectral method.

**Practical application** These algorithms in this category usually proposed for solving certain problems of graph partition in parallel graph computing, such as matrix computation, PageRank, triangle listing, etc. By analyzing the feature of the computing of the certain application, the researchers propose a graph partition algorithm. In this situation, the algorithm is simple but has good performance in the process of parallel graph computing. Erik G. Boman et al. [12] study a special case and propose a graph partition algorithm for scalable matrix computations on large scale-free graphs to reduce the running time by 2D graph partitioning. They test the algorithm on structurally symmetric matrices. Shumo Chu et al. propose [13] a simple graph partition algorithm for problem of triangle listing in massive networks. Semih Salihoglu et al. [14] propose an algorithm called *Large Adjacency-List Partitioning* in their graph processing system for PageRank. When they partition graph, they will store the vertices with large degree across the compute nodes. Actually, there will be some challenges in this algorithm, such as how to define the large degree, and what if there are too many vertices with large degree. The shortcoming of these algorithms for practical applications is that they may lack of rigorous theoretical analysis. But they are like object-oriented programming and usually perform very well for certain problem.

In this recent works, we can find that the algorithms in theoretical analysis are hardly applied to the parallel graph computing for kinds of limitations; for the practical application ones, one algorithm usually works for one special problem. Finding a general algorithm is very hard, because different problems on parallel graph computing have different requirements of the partition structure and the relationship between partitions. Even in some parallel graph systems, the graphs are partitioned randomly [4], which leads to poor performance. Up to now, no one has focused on the problem of random walks on partitions of parallel graph computing. In this paper, we will analyze this problem and propose a feasible graph partition framework for random walks. We also propose metrics to evaluate the result of the partitions.

## 7. CONCLUSION AND FUTURE WORK

In this work, we give detailed analysis for the process of random walks on partitions. Based on the analysis, two optimization functions are presented to reduce the number of times of communication and replications of vertices in the condition that the balance of the partitions can be ensured for parallel computing. A feasible graph partition framework for random walks is proposed by surveying the two optimization functions. The framework are the foundations of the four greedy graph partition algorithms. We also compare these algorithms and give the advantages and shortcomings of them. In order to measure the partition result, we also give five metrics from different views. The results of the algorithm on the graph of real world show the superior performance of our algorithms and metrics. We also design three tables to represent the partitions, which benefits the implementation of random walks on the partitions by parallel computing.

Here, we mainly introduce our idea of graph partition for random walks. There exist lots of works to be done. Although the preliminary results verify the correction of our analysis, more work will be followed up. More algorithms can be proposed in the framework to improve these ones. More graph data should be tested by these algorithms. How to determine the balance between the cost of the memory and storage and the cost of the bandwidth is also difficult to deal with.